\documentstyle[twocolumn,aps,epsf]{revtex}

\begin{document}
\draft
\title{All-Optical Production of a Degenerate Fermi Gas}
\author{S. R. Granade, M. E. Gehm,  K. M. O'Hara, and J. E. Thomas}
\address{Physics Department, Duke University,
Durham, North Carolina 27708-0305}
\date{\today}
\wideabs{\maketitle
\begin{abstract}
We achieve degeneracy in a mixture of the two lowest hyperfine
states of $^6$Li by direct evaporation in a CO$_2$ laser trap,
yielding the first all-optically produced degenerate Fermi gas.
More than $10^5$ atoms are confined at temperatures below
$4\,\mu$K at full trap depth, where the Fermi temperature for each
state is $8\,\mu$K. This degenerate two-component mixture is ideal
for exploring mechanisms of  superconductivity ranging from Cooper
pairing to Bose condensation of strongly bound pairs.
 \end{abstract}
\vspace*{.125in}
}

Degenerate two-component Fermi gases offer tantalizing
possibilities for precision studies of pairing interactions in
systems for which the density, temperature, and interaction
strength are widely variable. Of particular interest are certain
two-component mixtures of $^{40}$K and $^6$Li which exhibit
magnetically tunable Feshbach resonances, enabling variation of
the s-wave scattering interaction from strongly repulsive to
strongly attractive. Attractive mixtures in these systems are
analogs of superconductors, since they have been predicted to
undergo a superfluid transition as a result of Cooper pairing at
experimentally accessible temperatures~\cite{Superfluidity,Bohn}.
Recently, two groups have predicted the possibility of
superfluidity arising from strong pairing in the vicinity of the
Feshbach resonance~\cite{Holland,Timmermans}. Transition
temperatures of up to half the Fermi temperature are predicted to
result from the strong coupling of the two-state Fermi gas to the
bosonic molecular state which causes the resonance. Since most
high temperature superconductors achieve transition temperatures
of only a few percent of the Fermi temperature, two-state Fermi
gases may be the highest temperature Fermi superfluids ever
studied~\cite{Combescot}. Further, these systems may permit
observation of the transition from weak Bardeen-Cooper-Schrieffer
superfluidity to Bose condensation of strongly-bound
pairs~\cite{Randeria}.

In contrast to Bose-Einstein condensates, which can be prepared
and studied in magnetic traps, two-component Fermi superfluids
must be prepared in state-independent optical dipole traps, since
the required pairs of hyperfine states in $^6$Li and $^{40}$K are
high-field seeking~\cite{Superfluidity,Bohn,O'Hara2}.  A
degenerate Fermi gas has been produced by direct evaporation of a
two-state mixture of $^{40}$K in a magnetic trap, using a dual
radio-frequency-knife method~\cite{Jin}. Sympathetic cooling of
fermionic $^6$Li to degeneracy also has been achieved by using
mixtures of $^6$Li with bosonic $^7$Li  in a magnetic
trap~\cite{Hulet,Salomon}. However, to explore superfluidity in
these systems,  transfer to an optical trap and subsequent state
preparation is required. The procedure for preparing an optically
trapped two-state Fermi gas can be greatly simplified by direct
evaporation in an optical trap.

In this Letter, we demonstrate all-optical production of a
degenerate mixture of the two lowest hyperfine states of fermionic
$^6$Li  in a stable, CO$_2$ laser trap~\cite{O'Hara1}. The trap is
loaded from a magneto-optical trap (MOT) at an initial temperature
of $150\,\mu$K. Degeneracy is obtained by forced evaporation,
accomplished by continuously lowering the trap depth; the trap is
then adiabatically recompressed to full depth. At this stage, more
than $10^5$ atoms remain at temperatures below $4\,\mu$K, less
than half of the Fermi temperature of $8\,\mu$K. These results are
consistent with scaling laws we have derived for the phase-space
density as a function of trap depth~\cite{Scaling}.

Our $^6$Li experiments employ a CO$_2$ laser trap with a single
focused beam, rather than a crossed-beam geometry as used recently
to produce a $^{87}$Rb Bose-Einstein condenstate (BEC) by forced
evaporation~\cite{Chapman}. Nevertheless, after free evaporation
at full trap depth, we  achieve a very high initial phase-space
density of $\simeq 8\times 10^{-3}$, somewhat larger than that
obtained after free evaporation in the $^{87}$Rb BEC experiments.

A commercial, radio-frequency-excited  CO$_2$ laser (Coherent-DEOS
LC100-NV) provides 140 W at $\lambda =10.6\,\mu$m for the trapping
laser beam. An Agilent (6573A) power supply produces stable
current for the radio-frequency source, yielding a very stable
laser intensity. The laser output is deflected by an acousto-optic
(A/O) modulator to control the power. A cylindrical ZnSe telescope
corrects the output of the A/O for ellipticity, and the beam is
expanded by a factor of 10 before passing through an aspherical
19.5 cm focal length lens. This lens focuses the beam into the
vacuum system, yielding a $1/e^2$ intensity radius of 47 $\mu$m.
The corresponding Rayleigh length is $z_0=660\,\mu$m. With an
incident power of 65 W in the trap region, the trap depth is
estimated to be 690 $\mu$K. The corresponding radial (axial)
oscillation frequency for $^6$Li is predicted to be 6.6 kHz (340
Hz), with a geometric mean of $\nu =(\nu_x\nu_y\nu_z)^{1/3}=2400$
Hz.

 The radial oscillation frequency is measured by modulating
the frequency of the A/O to produce a sinusoidal displacement at
the trap focus with an amplitude of 0.2 $\mu$m. After the sample
is initially prepared at a temperature of $\simeq 15\,\mu$K, the
modulation is applied  for one second. The number of remaining
atoms is measured by resonance fluorescence. Repeating this
procedure as a function of modulation frequency reveals a
resonance in the trap loss at 6.5 kHz, in close agreement with
predictions. Parametric resonance methods~\cite{Hansch} yield
results consistent with the expected radial and axial oscillation
frequencies after correction for the expected resonance frequency
shift~\cite{Parametricmodel}.

Extremely low residual heating rates are attained in the
experiments.  At the maximum trap intensity of $1.9$ MW/cm$^2$,
the optical scattering rate is  2 photons per hour as a
consequence of the 10.6 $\mu$m wavelength~\cite{Knize}, yielding a
recoil heating rate of only 16 pK/sec. At the  background pressure
of $<10^{-11}$ Torr, heating arising from diffractive background
gas collisions~\cite{Bali,Beijerinck} is $<5$ nK/sec. For the trap
radial oscillation frequency of 6.6 kHz, the intensity noise
heating time constant is estimated to be $>2.3\times 10^4$ seconds
based on the measured laser intensity noise power
spectrum~\cite{O'Hara1}. A residual heating rate $<5$ nK/sec is
measured at full trap depth over 200 seconds. Trap 1/e lifetimes
of 400 seconds are observed.

The CO$_2$ laser trap is continuously loaded from a $^6$Li MOT.
The MOT is loaded from a Zeeman slower for 5 seconds, after which
the MOT laser beams are tuned $\simeq 6$ MHz below resonance and
lowered in intensity to $0.1\,I_{sat}=0.25\,{\rm mW}/{\rm cm}^2$
to obtain a Doppler-limited temperature of $\simeq 150\, \mu$K at
a density of $10^{11}$/cm$^3$. Following this loading stage, the
MOT gradient magnets are extinguished and the upper $F=3/2$
hyperfine state is emptied to produce a 50-50 mixture of atoms in
the lower $|F=1/2,M=\pm 1/2\rangle$ states~\cite{O'Hara2}.

The $|F=1/2,M=\pm 1/2\rangle$ mixture is of particular interest,
as it is predicted to  exhibit a Feshbach resonance near 850
G~\cite{Elastic}. A convenient feature of this mixture is that the
s-wave scattering length vanishes in the absence of a bias
magnetic field~\cite{Elastic}. However, the scattering length
varies between 0 and $-300\,a_0$ as the bias magnetic field is
tuned between 0 and 300 G~\cite{Elastic}. Hence, rapid evaporation
can be turned on and off simply by applying or not applying a bias
magnetic field.

The number of  trapped atoms is enhanced by increasing the
intensity of the CO$_2$ laser during the loading
stage~\cite{Loading}. To accomplish this, the beam which emerges
from the trap is recollimated after a ZnSe exit window by a 19.5
cm focal length ZnSe lens, and then retroreflected and
orthogonally polarized using a rooftop mirror oriented at
$45^\circ$ to the incoming polarization. The resulting
backward-propagating beam  is refocused into the trap region
through the exit lens. After passing through the trap region, this
beam is diverted by a thin film polarizer to a beam dump to avoid
feedback into the laser. Typically $1.5\times 10^6$ atoms are
confined in the forward propagating trap beam alone. The
backward-propagating beam increases this number to $3.5\times
10^6$.

After the CO$_2$ laser trap is loaded, the atoms are precooled by
free evaporation. To initiate evaporative cooling, we apply a bias
magnetic field of 130 G by reversing the current in one of the MOT
gradient coils, yielding a scattering length of $\simeq
-100\,a_0$. During free evaporation, a pneumatically controlled
mirror slowly blocks the backward-propagating beam by  diverting
the power into a 100 W power meter. Since this beam is refocused,
the trap region is  Fourier-transform-related to the plane of the
blocking mirror, and the trap smoothly evolves into a single beam
configuration. After 6 seconds of free evaporation, the single
beam trap contains $N=1.3\times 10^6$ atoms  at a temperature
 $T=50\,\mu$K. This precooling procedure provides excellent
initial conditions for the forced evaporation experiments, since
the resulting phase-space density for each state at full trap
depth, $\rho_i=(N/2)(h\nu)^3/(k_BT)^3$, is $8\times 10^{-3}$,
which is extremely high.

In all of our experiments, we characterize the velocity
distribution of the trapped gas by time-of-flight imaging. We use
the A/O modulator to turn off the CO$_2$ laser trap abruptly
($\Delta t<1\,\mu$s), permitting the gas to expand for a time
between $400\,\mu$s and 1.2 ms in zero bias magnetic field.
Residual A/O leakage is reduced to less than $2\times 10^{-5}$ of
the maximum intensity by extinguishing the radio-frequency
synthesizer output prior to the amplifier. Then a linearly
polarized probe laser pulse with a resonant intensity of
$0.1\,I_{sat}$ and a detuning of 3 half linewidths ($\simeq 9$
MHz) illuminates the gas for $10\,\mu$s. Simultaneously, a
noncopropagating repumper, resonant with the D2 lines starting
from the $F=3/2$ state, suppresses optical pumping into the upper
$F=3/2$ hyperfine state. The probe detuning reduces sensitivity to
the unresolved excited state hyperfine structure and light shifts
from the resonant repumper. For the selected 9 MHz detuning, the
expansion time is chosen so that the imaged cloud has a small
optical absorption $<35$\%.  An achromat at the vacuum system exit
window produces a 1:1 image of the atomic distribution in an
intermediate plane. This plane is imaged onto a CCD camera (Andor)
using a microscope objective to produce a net magnification of
$\simeq 4$. The magnification is calibrated by moving the axial
position of the trap focus through $\pm 1.25$ mm using a
micrometer-controlled translation stage. Fitting the central peak
of the distribution to a straight line yields a magnification of
$3.9$.

The images are processed to obtain the transverse spatial
distribution by integrating the measured optical depth in the
axial direction. In typical measurements, the cloud expands
ballistically by $100-200\,\mu$m  in $400\,\mu$s, much larger than
its initial transverse dimension. In the classical regime, we
assume ballistic expansion with a Maxwellian distribution. In this
case, the temperature is readily determined from the transverse
1/e width of the cloud: $a(t)=v\sqrt{1/(2\pi\nu_r)^2+t^2}$, where
$v=\sqrt{2k_BT/M}$ is the thermal velocity and $t$ is the time.
Since $\nu_r=6.6$ kHz, for $t>>24\,\mu$s, $a(t)=vt$. Measurements
of the cloud radius for several expansion times between
$100\,\mu$s and $600\,\mu$s fit very well to a straight line.

The number of atoms is determined from the  spatially integrated
optical depth of the absorption image and the absorption cross
section $\sigma$. For each of the $M=\pm 1/2$ magnetic sublevels
of the populated $F=1/2$ state,  $\sigma$  is  taken to be
$(\lambda^2/\pi)/(1+(2\Delta/\gamma)^2 )$, $2/3$ of that of the
cycling transition. Since the excited  hyperfine states are
unresolved compared to the linewidth $\gamma =5.9$ MHz, this cross
section contains contributions from both allowed transitions.
Results for the number are consistent within 10\% for several
detunings $\Delta$ between $9$ to $30$ MHz and $-30$ to $-9$ MHz,
and for variation of the camera focal plane over $\pm 1$ mm from
the plane which gives the sharpest image.

After precooling by free evaporation,  further cooling is
accomplished by lowering the trap depth, producing forced
evaporation. We have developed scaling laws for the number of
atoms, collision rate, and phase-space density as a function of
trap depth $U$ for an optical trap which is continuously
lowered~\cite{Scaling}. These scaling laws are valid for a fixed
$\eta =U/(k_BT)>>1$. For $\eta =10$, we find that the ratio of the
final to initial phase-space density increases according to
$\rho/\rho_i=(U_i/U)^{1.3}$.   This result shows that lowering the
trap depth by a factor of 100 should increase the phase-space
density by a factor of 400, producing a degenerate sample for
$\rho_i>2.5\times 10^{-3}$. To maintain a constant value of
$\eta$, the trap should be lowered from its initial depth $U_i$
according to the formula
\begin{equation}
U(t)=U_i/(1+t/\tau )^\beta,
 \label{eq:trapdepth}
  \end{equation}
  which assures that the lowering rate slows as the collision rate
  decreases~\cite{Scaling}. Taking $\eta =10$, we have $\beta =1.45$ and
$1/\tau =2.0\times 10^{-3}\,\gamma_i$, where $\gamma_i$ is the
initial elastic collision rate. For a 50-50 mixture of fermions,
$\gamma_i =\pi N_iM\sigma\nu_i^3/(k_BT_i)$ with $N_i$ the initial
total number of atoms. Note that $\gamma_i$ is reduced by a net
factor of 4 compared to a single-component Bose gas with the same
parameters.  For a scattering length of $a\simeq -100\,a_0$, the
elastic  cross section is $\sigma =8\pi a^2=0.7\times
10^{-11}\,{\rm cm}^2$. Using $\nu_i=(\nu_x\nu_y\nu_z)^{1/3}=2.4$
kHz, $N_i=1.0\times 10^6$, and $T_i=50\,\mu$K, we obtain $\gamma_i
=4.4\times 10^2\,{\rm s}^{-1}$ and $\tau =1.1$ sec.

Unfortunately, the A/O modulator that controls the CO$_2$ laser
intensity produces an ellipticity which varies as the
radio-frequency (rf) power is varied. The ellipticity is corrected
at maximum rf power by a cylindrical telescope. However, the
telescope provides only fixed compensation. Hence, as the rf power
is decreased to lower the trap depth, the beam becomes elliptical,
reducing $\nu_x\nu_y\nu_z$ by a factor of two compared to that
expected on the basis of the laser power alone. Further, we find
that the direction of the beam changes by 3 mrad as the rf power
is reduced by a factor of 100, causing vignetting. We align the
trap beam to  minimize this vignetting, but beam distortion still
occurs. For this reason, we cannot accurately compare our
evaporation results to the scaling law model. To compensate for
the  loss of confinement arising from the beam distortion as the
trap is lowered, we increase $\tau$ to 3 seconds. The trap laser
intensity is lowered using  an Agilent (33120A) arbitrary waveform
generator, the output of which is filtered with  a time constant
of 0.2 sec before being applied to the multiplier input of the A/O
radio-frequency generator.
\begin{figure}
\begin{center}\
\epsfysize=60mm \epsfbox{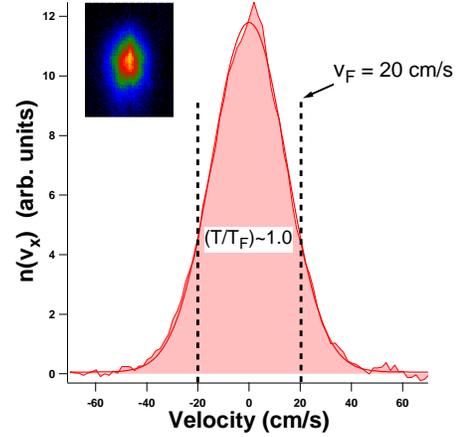}
\end{center}
 \caption{Absorption image (inset) and velocity distribution after 10
seconds of  forced evaporative cooling followed by recompression
to full trap depth. An average of 5 trials is shown. $T/T_F=1$, as
determined by a fit to a Maxwellian distribution. $v_F=20$ cm/s is
the Fermi velocity for a total $N=8\times 10^5$ atoms.}
\label{fig:1}
\end{figure}
We have measured atomic  velocity distributions  after forced
evaporative cooling for a variable time $t_f$. To provide a
calibrated reference trap, time-of-flight images are recorded
after adiabatic recompression to full trap depth over 11 seconds.
This also increases the spatial density and hence the Fermi
temperature, while preserving the phase-space density.
Fig.~\ref{fig:1} shows the velocity distribution for $t_f= 10$
seconds. The total number of atoms remaining is $N=8\times 10^5$,
corresponding to a Fermi temperature of $T_F=h\nu
(6N/2)^{1/3}/k_B=15\,\mu$K  for each state.  Assuming a Maxwellian
distribution, the gas is at a temperature of $15\,\mu$K, yielding
$T/T_F =1$. At this temperature, a substantial number of atoms
have velocities greater than the Fermi velocity of 20 cm/sec.

 Near degeneracy, the energy of the atoms
contains a contribution from the Fermi energy so that the true
temperature is lower than that obtained using a Maxwell-Boltzmann
(MB) distribution which assumes that all of the energy is thermal.
Hence, the low temperature absorption images are fit using a
Thomas-Fermi (TF) approximation to determine
$T/T_F$~\cite{Butts,DeMarco}, where the Fermi temperature $T_F$ is
calculated using the measured trap frequencies and integrated atom
number. At the lowest temperatures achieved in the experiments,
the MB temperature is $\simeq 10$\% higher than the TF
approximation.

Degeneracy is attained for $t_f=40$ seconds, where $T\simeq
5.8\,\mu$K and $T/T_F=0.55$ with $3\times 10^5$ atoms remaining.
At this temperature, the gas is degenerate, and $\rho\simeq
(T_F/T)^3/6\simeq 1$~\cite{Butts}. We have also measured the
temperature of the atoms in the lowered trap without recompression
to full trap depth. We obtain temperatures a factor of $\simeq 10$
lower, i.e., $\simeq 580$ nK, as expected for a harmonic trap
which is lower in depth by a factor of $\simeq 100$.

\begin{figure}
\begin{center}\
\epsfysize=60mm \epsfbox{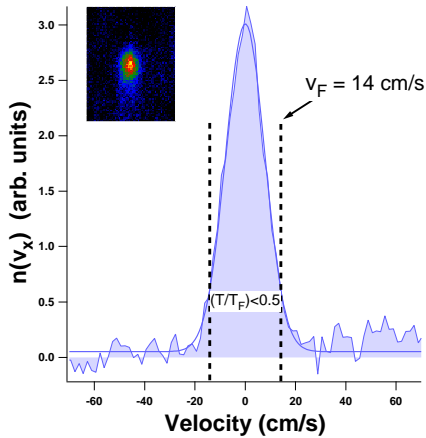}
\end{center}
\caption{Absorption image (inset) and velocity distribution after 60
seconds of  forced evaporative cooling followed by recompression
to full trap depth. An average of 5 trials is shown.  $T/T_F<0.5$
as determined by a fit using a Thomas-Fermi approximation.
$v_F=14$ cm/s is the Fermi velocity for a total $N=10^5$ atoms.}
\label{fig:2}
\end{figure}

 Fig.~\ref{fig:2} shows the velocity distribution for $t_f=60$ seconds.
  The total number of atoms is reduced
to $10^5$, corresponding to a Fermi temperature of $8\,\mu$K. The
measured temperature is below $4\,\mu$K, yielding $T/T_F =0.48$.
Nearly all atoms have velocities less then the Fermi velocity of
14 cm/sec.

In the experiments, we achieve high evaporation efficiency
$\chi\equiv \ln (\rho_f/\rho_i)/\ln (N_i/N_f)$~\cite{Efficiency}.
 For example,
after precooling, but prior to forced evaporation, $N_i=1.3\times
10^6$ and $\rho_i= 8\times 10^{-3}$ per state. After 40 seconds of
forced evaporation, $N_f=0.3\times 10^6$ and $\rho_f\simeq 1$.
Hence, $\chi \simeq 3.3$. The overall evaporation efficiency is
similar. Starting with the loading conditions where the total
number of atoms is $N_i=3.5\times 10^6$ at a temperature of
$150\,\mu$K, we obtain $\chi =2.9$ after 40 seconds of forced
evaporation. Despite the trap distortion described above, these
results are comparable to the best achieved in magnetic
traps~\cite{Efficiency}.

In conclusion, we have produced a degenerate, two-component $^6$Li
Fermi gas in a single beam all-optical trap by direct evaporative
cooling. By using a stable CO$_2$ laser trap at a background
pressure of $<10^{-11}$ Torr, efficient evaporation over time
scales of 85 seconds is achieved. In future experiments, it will
be possible to attain scattering lengths of $\simeq -300\,a_0$ by
increasing the bias magnetic field to 300 G, thereby increasing
the elastic cross section at low temperature by nearly factor of
10. This should enable preparation of a degenerate sample in just
a few seconds, producing substantially lower temperatures by
reducing the detrimental effects of any residual heating. We are
currently preparing for a  systematic study of the Feshbach
resonance at higher magnetic field, and hope to observe superfluid
pairing in a two-state Fermi gas.

 This research is supported by
the Physics divisions of the Army Research Office and the National
Science Foundation, the Fundamental Physics in Microgravity
Research program of the National Aeronautics and Space
Administration, and the Chemical Sciences, Geosciences and
Biosciences Division of the Office of Basic Energy Sciences,
Office of Science, U. S. Department of Energy.

\end{document}